\documentclass{mem}
\usepackage{natbib}\usepackage{txfonts}\usepackage{balance}
\usepackage{graphicx}
\usepackage[a4paper]{hyperref}
\idline{00}{1}
\begin{document}
\def\teff{$T\rm_{eff }$}
\def\kms{$\mathrm {km s}^{-1}$}

\title{Angular Diameter Amplitudes of Bright Cepheids}

\subtitle{}

\author{P.\thinspace Moskalik\inst{1} \and N.\thinspace A.\thinspace Gorynya\inst{2}}

\offprints{P. Moskalik, \email{pam@camk.edu.pl}}

\institute{
Copernicus Astronomical Centre,
ul. Bartycka 18, 00--716 Warsaw, Poland
\and
Institute of Astronomy, Russian Academy of Sciences,
48 Pyatnitskaya St., 109017 Moscow, Russia
}

\authorrunning{Moskalik \& Gorynya }
\titlerunning{Angular Diameter Amplitudes}

\abstract{Expected mean angular diameters and amplitudes of angular diameter
          variations are estimated for all monoperiodic Classical Cepheids
          brighter than $\langle V \rangle = 8.0$\thinspace mag. The catalog
          is intended to help selecting best Cepheid targets for
          interferometric observations.

}

\maketitle{}

\section{Introduction}

Classical Cepheids play a central role in establishing
cosmological distance scale. An accurate calibration of their
$P-L$ relation is, therefore, crucial. While the slope of this
relation is well determined by Cepheids of the Large Magellanic
Cloud, its zero point still remains uncertain. The long-baseline
optical interferometry offers a novel way of Cepheid distance
determination, by using purely geometrical version of the
Baade-Wesselink method ({\it e.g.} Kervella et al. 2004). So
far, the technique was successfully applied only to a handful of
Cepheids, but with increased resolution of next generation
instruments (CHARA and AMBER) more stars will become accessible.

The goal of this work is to identify Cepheids, which are most
promising targets for observations with existing and future
interferometers. For that purpose, we calculated expected mean
angular diameters and angular diameter amplitudes for all
monoperiodic Cepheids brighter than $\langle V \rangle =
8.0$\thinspace mag. The resulting catalog can serve as a planning
tool for future interferometric observations. Full version of the
catalog can be found in Moskalik \& Gorynya (2005; hereafter
MG05).

\begin{table}
\caption{Predited Angular Diameters of Bright Cepheids (for full table see MG05)}
\vskip -0.2cm
\label{tab1}
\begin{center}
\begin{tabular}{lcccc}
\hline
\noalign{\smallskip}
Star          & $\log P$
                      & $\langle V\rangle$
                              & $\langle\theta\rangle$
                                      & $\Delta\theta$ \\
\noalign{\smallskip}
\hline
\noalign{\smallskip}
 $\ell$ Car   & 1.551 & 3.724 & 2.854 & 0.545 \cr
 SV Vul       & 1.653 & 7.220 & 1.099 & 0.270 \cr
 U Car        & 1.588 & 6.288 & 1.059 & 0.252 \cr
 RS Pup       & 1.617 & 6.947 & 1.015 & 0.250 \cr
 $\eta$ Aql   & 0.856 & 3.897 & 1.845 & 0.226 \cr
 T Mon        & 1.432 & 6.124 & 0.949 & 0.219 \cr
 $\beta$ Dor  & 0.993 & 3.731 & 1.810 & 0.214 \cr
 X Cyg        & 1.214 & 6.391 & 0.855 & 0.184 \cr
 $\delta$ Cep & 0.730 & 3.954 & 1.554 & 0.181 \cr
 RZ Vel       & 1.310 & 7.079 & 0.699 & 0.170 \cr
 $\zeta$ Gem  & 1.006 & 3.918 & 1.607 & 0.160 \cr
 TT Aql       & 1.138 & 7.141 & 0.800 & 0.158 \cr
 W Sgr        & 0.881 & 4.668 & 1.235 & 0.151 \cr
\noalign{\smallskip}
\hline
\end{tabular}
\end{center}
\end{table}

\section{Method}

First, Cepheid absolute magnitudes were estimated with the
period--luminosity relation of Fouqu\'e et~al. (2003). The
observed periods of first overtone Cepheids were fundamentalized
with the empirical formula of Alcock et~al. (1995). Comparison of
derived absolute magnitudes and dereddened observed magnitudes
yielded Cepheid distances.

The mean Cepheid radii were estimated with the period--radius
relation of Gieren et~al. (1998). Variations of Cepheid radii
during pulsation cycle were calculated by integrating the observed
radial velocity curves. For all Cepheids we used the same constant
projection factor of $p=1.36$. With the mean radius, the radius
variation and the distance to the star known, the mean angular
diameter, $\langle\theta\rangle$, and the total range of angular
diameter variation, $\Delta\theta$, can be easily calculated.

\section{Results}

Results of our calculations are summarized in Table\thinspace 1.
In Fig.\thinspace 1, we display $\langle\theta\rangle$ and
$\Delta\theta$ {\it vs.} pulsation period for all Cepheids of our
sample.

At the level of technology demonstrated already by VINCI/VLTI and
PTI instruments, the achievable accuracy of $\langle\theta\rangle$
determination is about 0.01\thinspace mas. This implies a lower
limit of $\langle\theta\rangle = 1.0$\thinspace mas, if
measurement with 1\% accuracy is required. Angular diameters of 13
Cepheids are above this limit, four of which have not been yet
observed (SV~Vul, U~Car, RS~Pup and overtone pulsator FF~Aql).

Most interesting for interferometric observations are Cepheids,
whose angular diameter {\it variations} can be detected. This has
been possible for stars with $\Delta\theta > 0.15$\thinspace mas.
13 Cepheids are above this limit. These objects cover uniformly
the period range of $\log P = 0.73 - 1.65$ and are well suited for
calibration of Cepheid $P-L$ and $P-R$ relations. Until now,
angular diameter variations have been measured only for six of
them. The remaining seven, so far unobserved Cepheids, are
SV\thinspace Vul, U\thinspace Car, RS\thinspace Pup, T\thinspace
Mon, X\thinspace Cyg, RZ\thinspace Vel, and TT\thinspace Aql. We
encourage observers to concentrate their efforts on these objects.

With shorter wavelength ($H$-band instead of $K$-band) and longer
baselines, the new CHARA and AMBER interferometers will offer
substantial increase of resolution.  Consequently, the list of
Cepheids with measurable amplitude diameter variations will grow
to $\sim$\thinspace 30 objects, creating excellent prospect for
very accurate calibration of Cepheid $P-L$ and $P-R$ relations.

\begin{figure}[]
\resizebox*{\hsize}{!}{\includegraphics[clip=true]{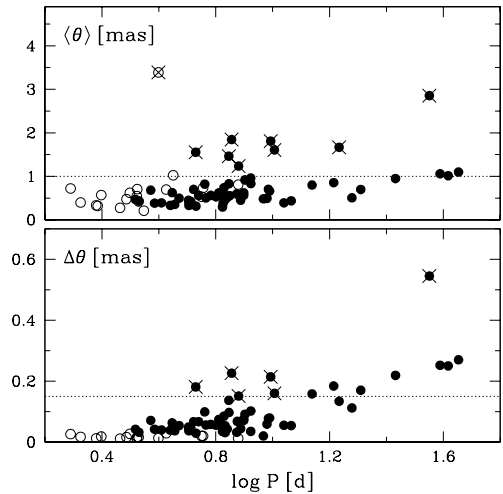}}
\caption{\footnotesize Predicted mean angular diameters (top) and
         angular diameter amplitudes (bottom) for Pop.\thinspace I \ Cepheids
         with $\langle V\rangle < 8.0$\thinspace mag. Fundamental and
         overtone pulsators plotted with filled and open circles,
         respectively. Stars with published interferometric measurements
         displayed with crossed symbols.}
\label{fig1}
\end{figure}

\bibliographystyle{aa}

\end{document}